\documentclass[letter]{aa} 
\usepackage{amsmath,mathrsfs}
\usepackage{graphicx}
\usepackage{placeins}
\usepackage[utf8]{inputenc}
\usepackage{filecontents}
\usepackage{xspace}
\usepackage{txfonts}
\usepackage{natbib}
\usepackage{multicol,multirow}
\newcommand{\Gaia}{\emph{Gaia}\xspace}

\newcommand{\Teff}{$T_{\rm eff}$\xspace}
\newcommand{\logg}{$log(g)$\xspace}
\newcommand{\feh}{[Fe/H]\xspace}
\newcommand{\mgfe}{[Mg/Fe]\xspace}
\newcommand{\sife}{[Si/Fe]\xspace}
\newcommand{\meh}{$[M/H]$\xspace}
\newcommand{\kms}{km~s$^{-1}$\xspace}

\newcommand{\energyunit}{km$^{2}$~s$^{-2}$\xspace}
\newcommand{\angmomunit}{km~s$^{-1}$~kpc\xspace}
\newcommand{\Lz}{$L_z$\xspace}

\begin{document}

\title{Flavours in the box of chocolates: chemical abundances of
  kinematic substructures in the nearby stellar halo}

\titlerunning{Chemical abundances of substructures in the nearby stellar halo}

\author{Jovan Veljanoski\inst{1} \and Amina Helmi\inst{1}}

   \institute{Kapteyn Astronomical Institute, University of Groningen,
              Landleven 12, 9747 AD Groningen, The Netherlands\\
              \email{jovan@astro.rug.nl}}

  \date{\today}


  \abstract
  {Stellar halos contain tracers of the assembly history
  of massive galaxies like our own. Exploiting the synergy between the TGAS and
  the spectroscopic RAVE surveys, \citet{2017A&A...598A..58H} recently
  discovered several distinct substructures in the Solar neighbourhood, defined
  in integrals of motion space. Some of these substructures may be examples of
  the building blocks that built up the stellar halo.}
  {We analyse the chemical properties of stars in these
  substructures, with focus on their iron and $\alpha$-element abundances as
  provided by the RAVE survey chemical pipeline.}
  {We perform comparisons of the \feh and \mgfe distributions of the
  substructures to that of the entire halo sample defined in the
  TGAS$\times$RAVE dataset.}
  {We find that over half of the nine substructures have
  $\sigma_{\text{\feh}} \leq 0.3$~dex. Two of the substructures have
  $\sigma_{\text{\feh}} \leq 0.1$~dex, which makes them possible
  remnants of disrupted globular clusters.  As expected most substructures and
  the vast majority of our stellar halo sample are $\alpha$-enhanced. Only one
  substructure shows a [Mg/Fe] vs \feh abundance trend distinct from
  the rest of the halo stars in our sample.}
  {}

   \keywords{Galaxy: kinematics and dynamics -- Galaxy: halo -- Solar neighbourhood -- abundances}

   \maketitle
%

\section{Introduction}
\label{sec:intro}

In the current cosmological model, much of the mass in galaxies similar to the
Milky Way is assembled hierarchically: pre-galactic fragments coalesce to form
larger objects. A by-product  of this process is the formation of extended,
diffuse stellar halos surrounding their host galaxies. Stellar halos thus keep a
record of the accretion  and merger events that happened from the onset of the
assembly process in the  early Universe, all the way to the present day. Indeed,
wide-field Galactic  surveys such as SDSS and PanSTARS have revealed much
substructure in the form of spatially coherent stellar streams
\citep[e.g.][]{2006ApJ...642L.137B,2014MNRAS.443L..84B,2016MNRAS.463.1759B} in
the outskirts of the Milky Way. Their findings are consistent with the idea that
the  outer stellar halo is  predominantly, if not solely build via  mergers.

In contrast, little is known about the assembly process of the inner stellar
halo, and what fraction of it is composed by stellar streams. Using kinematic
and metallicity data of stars in the extended Solar neighbourhood,
\citet{2007Natur.450.1020C,2010ApJ...712..692C} found those less bound to be on
average on retrograde orbits and to be more metal-poor in comparison to inner
halo stars, which on average have slightly prograde motions. This led to the
idea of a two- component halo, where the inner parts would have predominantly
formed \emph{in situ} while the outer regions via accretion. This dual formation
path of the Galactic stellar halo is partly supported by hydrodynamical
simulations \citep[e.g.][]{2009ApJ...702.1058Z,2014MNRAS.439.3128T}, where the
relative importance of the two channels can vary strongly depending on the
cosmological history of the galaxy.

Conversely, if the stellar halo was fully built by accretion, models predict
that over 300 streams should cross the Solar neighbourhood
\citep{1999MNRAS.307..495H,2003MNRAS.339..834H}, and the first such streams were
discovered nearly two decades ago \citep{1999Natur.402...53H}.  Due to the
shorter dynamical time-scales, the inner stellar halo is spatially well mixed
however, causing the velocities of the stars in any substructure to become more
clustered. Thus, the halo's granularity can only be discerned in large samples
with accurate kinematics.

The launch of the \Gaia satellite puts us in an unprecedented position to study
the properties of stellar halo of the the Milky Way. The first \Gaia Data
Release (DR1) provides the  positions, parallaxes, proper motions, and mean
$G$-band magnitudes for over  2~million stars in common with the \emph{Tycho-II}
and \textsc{hipparcos} catalogues, in what is referred to as the \emph{Tycho-
Gaia} Astrometric Solution
\citep[TGAS,][]{2016A&A...595A...2G,2016A&A...595A...4L}.

Recently, \citet[][H17 hereafter]{2017A&A...598A..58H} exploited the synergy
between the TGAS and the spectroscopic RAVE surveys
\citep[DR5,][]{2017AJ....153...75K} to select a sample of over 1000 halo stars
using kinematics and metallicity criteria. Analysing the velocities of the stars
in this sample by means of a correlation function, H17 found that the stellar
halo near the Sun is consistent with being fully formed by accretion. In
addition, these authors identified 10 distinct, statistically significant
substructures in integrals of motion space.

These previously unknown substructures are the focus of this paper. Here we
analyse the chemical composition of their constituent stars using the abundances
available through the public RAVE~DR5 catalogue. With the goal of learning more
about their possible nature, this study is also a proof-of-concept of the
chemical labelling/tagging ideas put forward by \cite{2002ARA&A..40..487F}. For
instance, if some of the substructures are remnants of disrupted globular
clusters, their stars are expected to be $\alpha$-enhanced with a very tight,
even negligible spread in iron abundance.

This paper is structured as follows. In the next section we revisit the halo
sample and the integrals of motion substructures of H17 by taking advantage of
the improved parallaxes and metallicities derived by \citet{2017arXiv170704554M}
for the TGAS$\times$RAVE  dataset.  In Section~\ref{sec:analysis} we analyse the
abundances of the substructures, and summarise our findings in
Section~\ref{sec:discussion}.

\begin{figure}
\centering
\includegraphics[width=\textwidth/2]{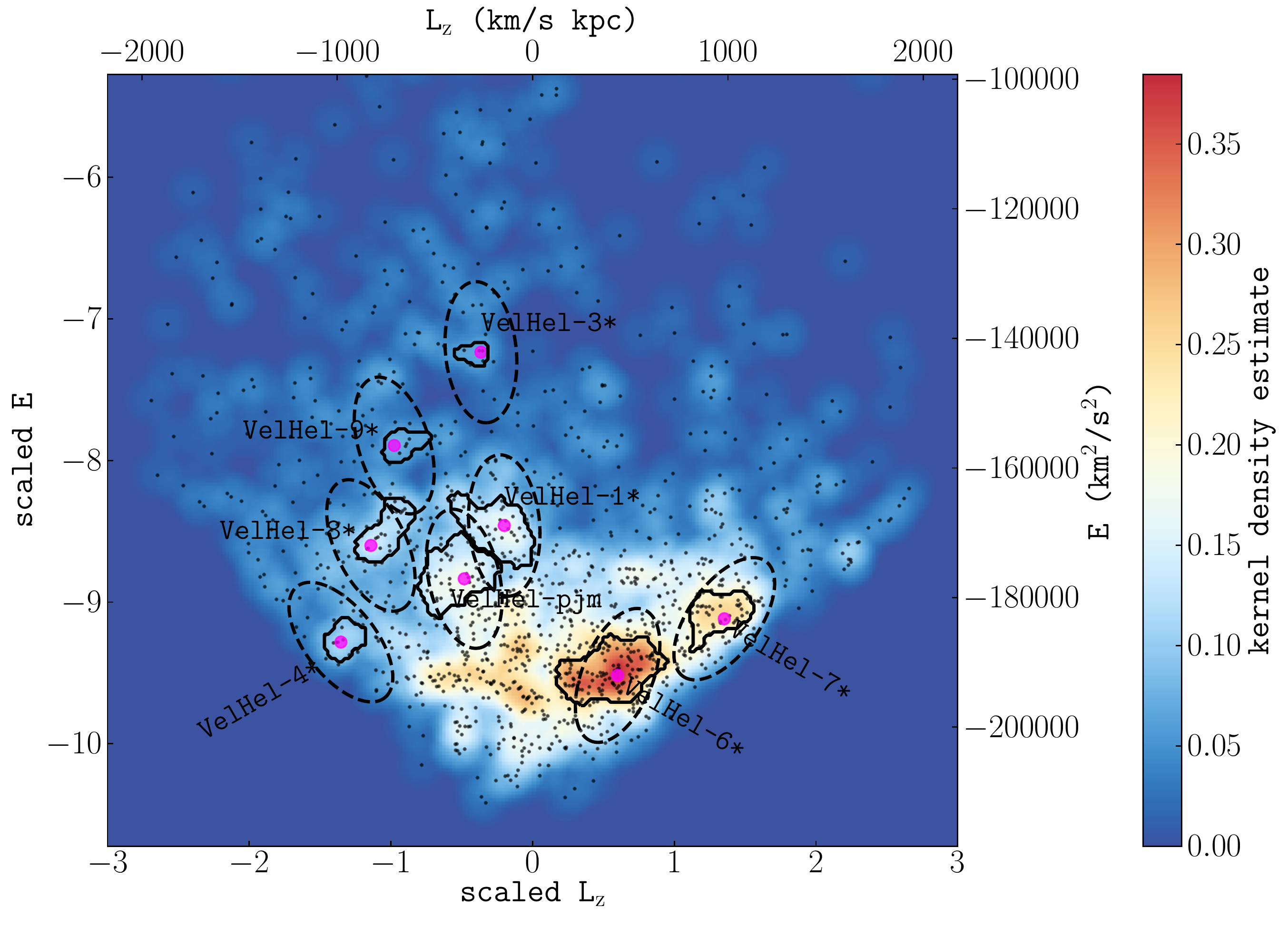}
\caption{Kernel density estimate of the distribution in $E-L_z$ for our sample
of halo stars in the Solar neighbourhood. The individual stars are shown as
black dots. The magenta points denote the statistically significant
over-densities. The black contours mark the extent of these  substructures,
as determined by the \texttt{watershed} algorithm. The ellipses show the
extended selection reported in Sec.~\ref{ss:ellipses}. The naming of the
structures follows the same convention as in H17.}
\label{fig:watershed}
\end{figure}

\section{``A box full of chocolates'' revisited}
\label{sec:remastered}

We start by redefining the halo substructures initially reported in H17, by
using the dataset constructed by \citet{2017arXiv170704554M}. These authors use
the TGAS parallaxes as priors when computing the spectrophotometric parallaxes
for stars in common between TGAS and RAVE. With this method the estimated
distances have uncertainties on average two times smaller
than their RAVE-only counterparts, and 1.4 times smaller than the
corresponding TGAS uncertainties. With their method \citet{2017arXiv170704554M}
also provide updated and more reliable \logg, \Teff and \meh estimates.

\begin{figure}
\centering
\includegraphics[width=\textwidth/2]{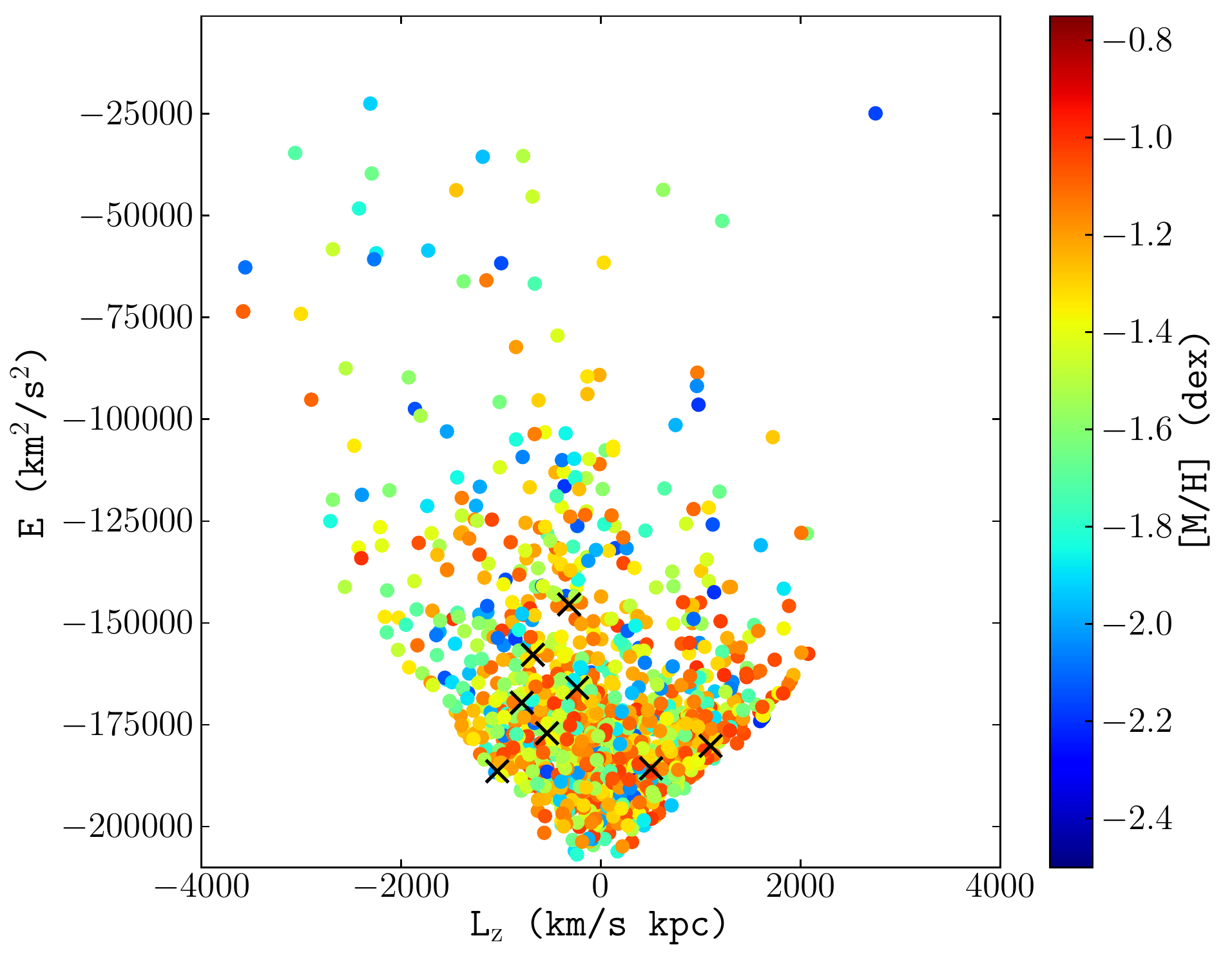}
\caption{Distribution of stars in our sample colour-coded by the RAVE
metallicity [M/H]. The black crosses indicate the location of the
overdensities identified in Fig.~\ref{fig:watershed}.}
\label{fig:mh}
\end{figure}

We define the sample of halo stars following the steps taken by H17 (see their
Sec.~2), using the TGAS$\times$RAVE catalogue provided by
\citet{2017arXiv170704554M}. After applying several quality
cuts\footnote{including a relative parallax uncertainties $\Delta\varpi/\varpi
\le 30\%$, where this can be the updated RAVE or TGAS-only parallaxes depending
on which measurement has a smaller relative error, provided that the TGAS
parallax is positive.}, and a metallicity criterion \meh$\le -1$~dex, we find
1823 tentative halo stars. In this work we use the ``calibrated'' metallicity
estimates provided by \citet{2017arXiv170704554M}. We then remove stars with
(thick) disk-like kinematics as in H17, by fitting two multivariate Gaussians,
and by retaining only those stars that have a higher probability of being drawn
from the Gaussian component with the lowest rotational motion (in our case, this
has a mean $v_y \approx -15$\kms). With this approach, we find 1217 high
confidence, metal-poor halo stars.

With this updated sample we proceed to re-analyse their distribution in
integrals of motion space: $E$-\Lz.  Fig.~\ref{fig:watershed} shows this
distribution for the more bound stars in  this halo sample (i.e. those that
satisfy $-21\times10^4 \le E \le   -10\times10^4$~\energyunit and $-2000 \le L_z
\le 2000$~\angmomunit).

To pick out any substructures that may be significant in this space, we proceed
again as in H17. We first determined the density field (shown in
Fig.~\ref{fig:watershed}), then applied a maximum filter to pick out
over-densities, and then established which were statistically significant by
comparison to randomised datasets obtained by reshuffling two velocity
components. In this way we identify 8 over-densities marked by the magenta
points on Fig.~\ref{fig:watershed}. Finally, we employ the \texttt{watershed}
algorithm \citep{watershed} to loosely determine the extent of the over-
densities and thus their constituent stars.

Fig.~\ref{fig:watershed}, shows that we have recovered all but one of the
substructures initially reported by H17 (see their Figure~10). Since not exactly
the same stars are found in the substructures as in H17 (there is typically 60\%
overlap), we added an asterisk at the end of their H17 labels. The one structure
that is not recovered, dubbed \texttt{VelHel-5}, can be visually identified in
our data, but does not pass the significance tests, although a new structure
appears that contains some stars from the original \texttt{VelHel-1} and
\texttt{VelHel-5}, which we label \texttt {VelHel-pjm}. We also do not recover
the metal-poor tail of the disk because, in contrast to H17,  we completely
removed the disk stars during the Gaussian decomposition phase.  Overall, the
above procedure adds more validity to the structures discovered by  H17, and
makes us appreciate how robust results can be to small changes in the  data.

In addition to the discoveries of the substructures in integrals of motion
space, H17 also found that the stars with binding energies smaller than that of
the Sun have preferentially retrograde orbits. We confirmed this result with our
updated halo sample: 77\% of the stars with $E >-13\times10^4$~\energyunit are
on retrograde orbits. We also include this low binding energy retrograde
component in our abundance analysis.
\begin{figure*}
\centering
\includegraphics[width=0.95\textwidth]{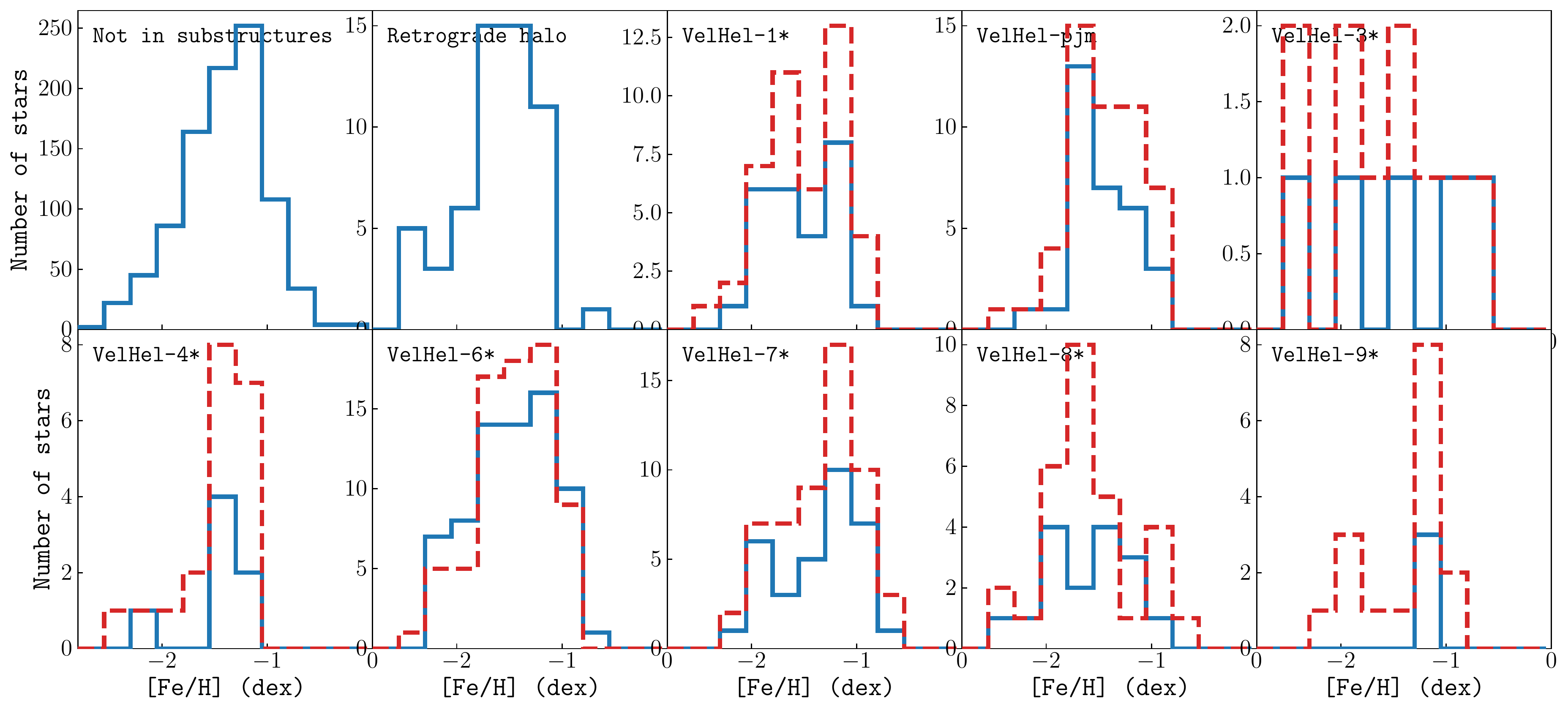}
\caption{The \feh distributions of the stars in the statistically
significant substructures identified in $E$-\Lz space, including the
retrograde halo. The blue solid histograms cover the stars identified using the
\texttt{watershed} algorithm, while those in red correspond to the extended
elliptical selection. The bins of all histograms have a constant width of
0.25~dex, which is the typical \feh uncertainty.}
\label{fig:fe_hist}
\end{figure*}

Fig.~\ref{fig:mh} shows the distribution of halo stars in the integrals of
motion space, colour-coded by the value of their metallicity. We can see that
the stars with high binding energy are, on average, more metal-rich, while those
in the less bound highly retrograde component appear to be somewhat more
metal-poor, albeit with large scatter. This would thus be in agreement with the
results of \citet{2007Natur.450.1020C}, although the ``outer halo'' component in
our dataset is much more prominently retrograde.

\section{Abundance analysis}
\label{sec:analysis}

The nominal RAVE~DR5 data release provides abundances of a few iron peak
elements (Fe, Ni), as well as several $\alpha$-elements (Mg, Si, Ti).  These
abundances were derived with a pipeline first introduced for the third data
release of RAVE \citep{2011AJ....142..193B}, and subsequently improved in later
releases. The typical abundance is $\sim 0.2$~dex, and can increase slightly
for $S/N < 40$. In addition, the RAVE~DR5 stars have a mean $S/N = 55$ and a
resolving power $R = \lambda/\Delta\lambda \sim 7500$.

Note that we do not use the abundance elements provided by \emph{The Cannon}
pipeline because the training set used is sparse in the regime of low
metallicity giants and hence not reliable.

\subsection{\feh and $\alpha$ abundances}
\label{sec:feh}

Figure~\ref{fig:fe_hist} shows the \feh distributions for the stars that
comprise the statistically significant substructures identified in
Section~\ref{sec:remastered}, as well as for the retrograde outer halo. One can
see that substructures \texttt{VelHel-4*}, \texttt{VelHel-pjm},
\texttt{VelHel-9*}, and even the retrograde halo have pronounced peaks in their
\feh distributions, and relatively small dispersions.

We now examine whether the small spread of \feh exhibited by the substructures
is genuine, which would separate them from the overall halo sample judging by
their iron abundances alone, or whether their observed \feh distributions are
simply chance occurrences. To quantify this we do the following test. We draw
10000 random sets of stars from the entire halo sample. Each random set contains
as many stars as the substructure that is being examined. We then count how
often a random set has a smaller \feh spread\footnote{What we refer to as
``spread'' is the difference between the values of the distribution at the 16th
and the  84th percentile, divided by two. This is equivalent to one standard
deviation in  the case of a Gaussian distribution.} compared to what is observed
in the actual substructures.

The results from this test are listed in Table~\ref{tab:stat}. From the second
column of this Table we see that 5 out the 9 substructures we are investigating,
including the retrograde halo, have \feh spreads of 0.3~dex or less. This is a
rather small value given the typical abundance uncertainties are 0.25~dex. For
substructures \texttt{VelHel-4*}, \texttt{VelHel-pjm*} and the retrograde halo,
we see that the probability of observing distributions with such small spread by
randomly selecting stars is $\sim 1\%$. Substructure \texttt{VelHel-9*} has the
smallest \feh spread amongst the substructures, but this quantity was measured
only from 3 stars, and such a configuration happens 6\% of the time in our
random realisations.

\begin{figure*}
\centering
\includegraphics[width=0.95\textwidth]{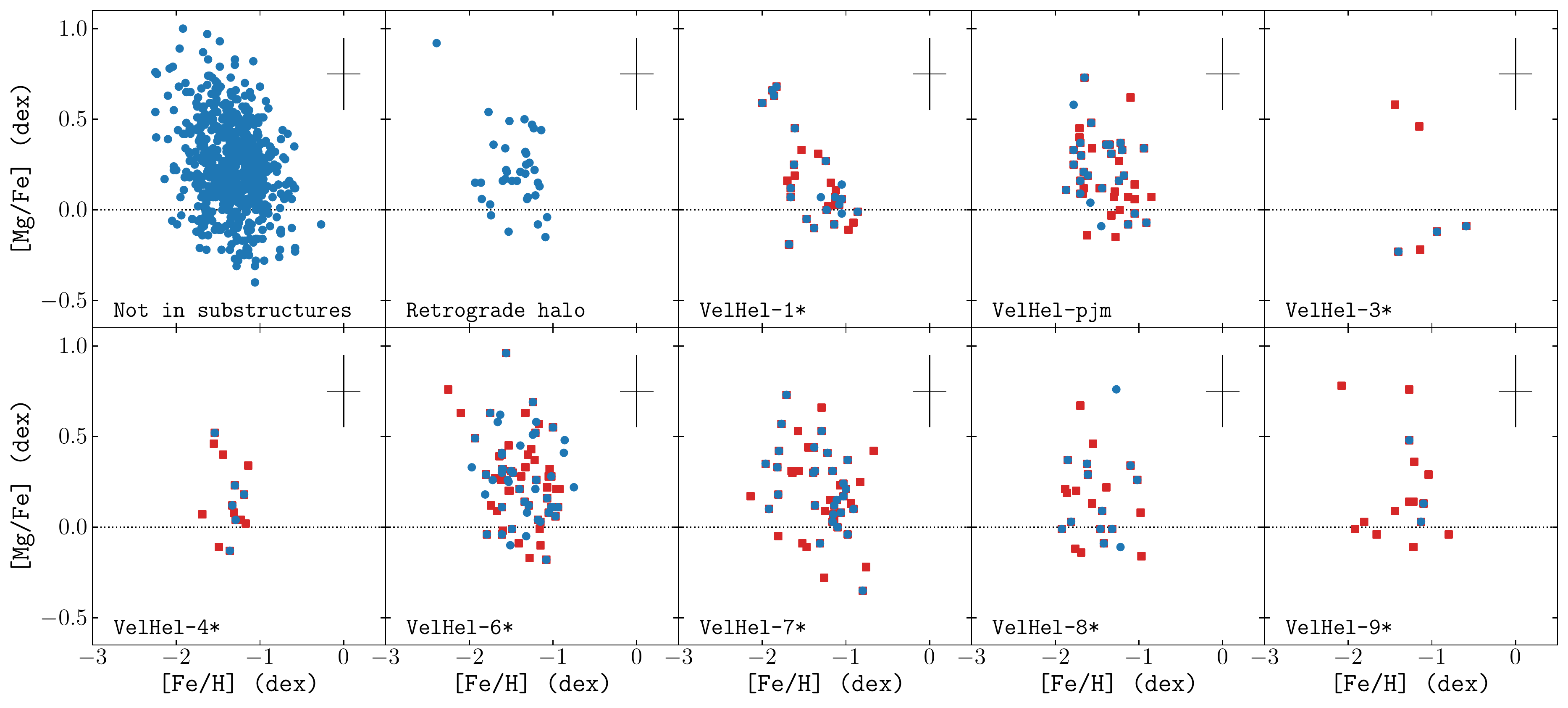} \\
\caption{\feh vs \mgfe (top) for the stars in the substructures, the retrograde
halo, and the rest of the halo sample. The colour-coding is the same as in
Fig.~\ref{fig:fe_hist}. One can see that the stars in most of the substructures
follow a similar distribution as most of the stellar halo sample (except for
those in \texttt{VelHel-3*}).}
\label{fig:fe_alpha}
\end{figure*}

Both Mg and Si are $\alpha$-elements and their abundance is available for the
largest fraction of stars that comprise our substructures. In
Fig.~\ref{fig:fe_alpha} we show the \mgfe as a function of \feh (\sife follows a
similar distribution).  One can see that almost all substructures appear to be
$\alpha$-enhanced. The only exception is substructure \texttt{VelHel-3*}, but
this result is still tentative at best, given that only 3 have Mg (and 4 stars
have Si) abundance estimates. This is intriguing \citep[given the ][low-$\alpha$
sequence]{2010A&A...511L..10N}, and a K-S test that compares the cumulative
distribution of the ``smooth'' stellar halo sample to that of
\texttt{VelHel-3*} confirms this: we find a $p-$value~$<0.01$.

\begin{table}
\centering
\caption{The standard deviations of the \feh  distributions of the $E$-\Lz
substructures and the less-bound retrograde halo. The table also shows the
frequency of observing a spread as small or smaller in the 10000 randomly drawn
sets from the full halo dataset (see main text for details).}
\label{tab:stat}
\begin{tabular}{lcccc}
\hline
\hline
                    & \multicolumn{2}{c}{Density selection} & \multicolumn{2}{c}{Elliptical selection} \\
\hline
Substructure        & $\sigma_{[Fe/H]}$                     & p$_{[Fe/H]}$                                & $\sigma_{[Fe/H]}$ & p$_{[Fe/H]}$ \\
                    & (dex)                                 & (\%)                                        & (dex)             & (\%)         \\
\hline
\texttt{VelHel-1*}  & 0.4                                   & 66                                          & 0.4               & 51          \\
\texttt{VelHel-3*}  & 0.7                                   & 99                                          & 0.5               & 90          \\
\texttt{VelHel-4*}  & 0.1                                   & $<1$                                        & 0.2               & $<1$        \\
\texttt{VelHel-6*}  & 0.4                                   & 28                                          & 0.3               & 1           \\
\texttt{VelHel-7*}  & 0.4                                   & 63                                          & 0.2               & $<1$        \\
\texttt{VelHel-8*}  & 0.3                                   & 44                                          & 0.4               & 61          \\
\texttt{VelHel-9*}  & 0.1                                   & 6                                           & 0.1               & $<1$        \\
\texttt{VelHel-pjm} & 0.2                                   & 2                                           & 0.3               & 2           \\
retr. halo          & 0.3                                   & 1.2                                         & ...               & ...         \\
\hline
\end{tabular}
\end{table}

\subsection{Enlarged membership selection}
\label{ss:ellipses}

One of the biggest challenges we are currently facing while trying to detect and
characterise spatially mixed substructures in the Solar neighbourhood is the
small number of halo stars that are available to us for analysis. In fact, the
main reason why examining integrals of motion diagrams is so beneficial, is
because we are effectively folding the 6-dimensional phase-space information
into two dimensions, thus increasing the clustering an object exhibits in an
$E$-\Lz space for example, as we show on Figure~\ref{fig:watershed}. As
described in Section~\ref{sec:remastered}, we detected the substructures
analysed in this paper as over-densities in $E$-\Lz space. The members of each
structure were determined with the \texttt{watershed} algorithm that traces the
density contour at some level of each structure. However, when the number of
members is small, the extent and shape of a structure in $E$-\Lz space is likely
not well defined. From simulations of spatially mixed substructures in the Solar
neighbourhood, we know that in $E$-\Lz space they have regular, almost
elliptical  shapes with a small dispersion in \Lz, while they can be quite
elongated in energy, depending on the orbit and mass of the progenitor system
\citep[see e.g. Figure~6 in][]{2000MNRAS.319..657H}. Thus it is possible that
other members of the substructures are present in the $E$-\Lz diagrams on
Fig.~\ref{fig:watershed}, but they are not selected simply due to the shape of
the underlying density field in the regime of low number statistics.

We thus attempt to analyse the elemental abundances of the substructures by
selecting all stars that fall within an ellipse centred on the peak of the over-
density identified in $E$-\Lz space, with a semi-major axis of 0.5 in scaled
units of energy, and a semi-major axis of 0.25 in scaled units of \Lz (since
$\Delta E/E \propto 2 \Delta L_z/L_z$), as shown in Fig.~\ref{fig:watershed}.
The major axes of the ellipses are aligned with the line connecting the density
peak of the substructures and the point of origin. Such a selection should
approximate the shape that a substructure is expected to have in $E$-\Lz in the
presence of significantly more members. We proceed to re- examine the Fe and
$\alpha$-element abundance distributions of the substructures when their members
are determined with this elliptical selection.

The red dashed histograms in Figure~\ref{fig:fe_hist} show that adding more
potential members to the substructures with the elliptical selection enhances
the features seen in the original \feh distributions. The most striking example
is substructure \texttt{VelHel-9*}, in which the tentative peak observed with
the nominal density selection is significantly enhanced: it now features 8 stars
with a dispersion smaller than 0.1~dex.

This simple exercise shows that, there are potentially several stars that are
likely members of the substructures which were not included via the original
\texttt{watershed}-based criteria. If this were not the case, we would expect
adding random stars to the substructures to smear out, and not improve, the
features seen in their \feh distributions. To quantify this, we repeat the
statistical test we conducted in Section~\ref{sec:feh}, now considering the
additional stars. The last two columns in Table~\ref{tab:stat} indeed show that
the increased prominence of the peaks makes the \feh distributions for most of
the structures less likely to be drawn by chance from the overall halo sample.
On the other hand, the elliptical membership selection does not significantly
change the distribution of stars in \mgfe vs \feh in our substructures, as shown
in Fig.~\ref{fig:fe_alpha}.

\section{Summary}
\label{sec:discussion}

In this paper the examined the iron and $\alpha$-element abundance distributions
of the substructures originally identified in integrals of motions space by
\citet{2017A&A...598A..58H} in the Solar neighbourhood, searching for hints of
their origin.

We started by defining a metal-poor halo sample following the steps described in
\citet{2017A&A...598A..58H} using the cross-match between the TGAS and RAVE
catalogues with the updated spectrophotometric distances provided by
\citet{2017arXiv170704554M}. With this procedure we selected 1217 high
confidence metal-poor halo stars. Using this data, we redefined the
substructures originally discovered by \citet{2017A&A...598A..58H} in $E$-\Lz
space. We recover all substructures (with at least 60\% overlap in terms of
membership) except for one, which is found to be less statistically significant.

We proceeded to analyse the Fe and Mg abundances available in RAVE~DR5 for stars
in the substructures, as well as for the retrograde halo stars that have $E>-13
\times 10^4$~\energyunit. These more loosely-bound retrograde stars are found to
be on average more metal-poor, in agreement with e.g.
\citet{2007Natur.450.1020C}.  We find that  over half of the identified
substructures have $\sigma_{\text{\feh}} \leq 0.3$~dex, which is rather small
given that the formal uncertainties of the individual star abundances in the
RAVE data are $\sim 0.25$~dex. In fact, substructures \texttt{VelHel-4*} and
\texttt{VelHel-9*} have $\sigma_{\text{\feh}} \leq 0.1$~dex, which makes them
good candidates for being remnants of disrupted globular clusters.

We find all substructures, save for \texttt{VelHel-3*}, to be $\alpha$-enhanced
in both Mg and Si as expected. In general, the $\alpha$
abundances of the stars in the substructures and in the retrograde halo
component are consistent with the rest of the stars in our halo sample. This is
also the case for their distribution in the space of [$\alpha$/Fe] vs \feh,
where the different substructures, with the exception of \texttt{VelHel-3*},
follow the same trend as the rest of the halo sample.

We are clearly limited in our ability to understand the structures by the low
number of stars with measured chemical abundances. This will change with the
\Gaia Data Release~2, that will certainly help us to better define the extent of
the substructures in integrals of motion space, and to find more members. But
more crucial are the upcoming spectroscopic surveys carried out by the new
multi-object spectrographs WEAVE and 4MOST, which will be of great help in
providing the data necessary to understand the nature and origin of these
objects.

\begin{acknowledgements} We gratefully acknowledge financial support from a VICI
grant from the Netherlands Organisation for Scientific Research, NWO and from
NOVA. This work has made use of data from the European Space Agency (ESA)
mission \Gaia (\url{http://www.cosmos.esa.int/gaia}), processed by the
\Gaia Data Processing and Analysis Consortium (DPAC,
\url{http://www.cosmos.esa.int/web/gaia/dpac/consortium}). Funding for the DPAC
has been provided by national institutions, in particular the institutions
participating in the \emph{Gaia} Multilateral Agreement.

This work made use of \texttt{vaex} \citep{2018arXiv180102638B}, \texttt{numpy}
\citep{Walt:2011:NAS:1957373.1957466}, \texttt{matplotlib} \citep{Hunter:2007},
\texttt{scikit-learn} \citep{2011JMLR.1953048.2078195}, and
\texttt{scikit-image} \citep{10.7717/peerj.453}.
\end{acknowledgements}

\bibliographystyle{aa}

\end{document}